\documentclass[12pt]{iopart}
\begin{document}
\title{A Background Independent Description of Physical Processes}
\author{M.\ Spaans}
\address{Kapteyn Institute, University of Groningen, 9700 AV Groningen, The Netherlands; spaans@astro.rug.nl}
\begin{abstract}
A mathematical structure is presented that allows one to define a
physical process independent of
any background. That is, it is possible, for a set of objects,
to choose an object from that set through a choice process that is defined
solely in terms of the objects in the set itself.
It is conjectured that this background free
structure is a necessary ingredient for a self-consistent description of
physical processes and that these same physical processes are
determined by the absence of any background.
The properties of the mathematical structure, denoted $Q$, are equivalent to
the three-dimensional topological manifold $2T^3\oplus 3(S^1\times S^2)$,
two three-tori plus three handles, embedded in four dimensions. The topology
of $Q$ reproduces the basic properties of QED and Einstein gravity.
\end{abstract}

\section{Introduction}

One of the outstanding questions in particle physics
is the unification of gravity with the electroweak and strong interactions.
Much effort has been devoted in the past decades to formulate a purely
geometrical and topological theory for these types of interactions.
Probably best known are the theories involving 11-dimensional super-gravity
and superstrings. These theories have been shown recently to be unified in
M-theory, although the precise formulation of the latter is not yet
known. A general problem in all this is the
formulation of a physical principle to lead the mathematical construction of
the theory. This paper attempts to contribute to such a principle, and is organized as follows.
Section 2 presents the guiding philosophy of this work, the no background principle.
Section 3 presents the resulting equations of motion and links them to QED and
Einstein gravity. Section 4 contains the discussion.
The Einstein summation convention is used.

\section{The No Background Principle}

One can ask the following question:
Is it possible, for a set of objects, to choose an object from that
set through a choice
process that is defined solely in terms of the objects in the set itself?
If such a construction can be found then it is self-contained in the sense that
it is independent of any background, like some fixed space-time in which objects interact.
Consequently, any theory that would derive from such a structure should be
internally self-consistent, and hopefully rich enough to be non-trivial.

Note then that, in order to choose an object $x$, one needs to {\it select} it and to
{\it distinguish} it from another object $y$.
Both these actions are required in order to assure that some notion of an
interaction between objects, like an observer using a measuring device or
a force between two particles, is at least possible.
In other words, one does not know, a priori, whether two selection actions
have not just singled out the same object (the process is blind).
With this definition of choice one can proceed to define selection and distinction operations
through topological constructs, the motivation being that topology naturally allows one to
work with continuous, but still discrete, building blocks, which is intuitively appealing.

First, the intersection of two loops, $S^1\times S^1$, facilitates selection of $x$.
Clearly, right from the start, the dimensionality of the space one works in is crucial.
Three spatial dimensions are very natural here since the two loops can be non-trivially linked,
so that they form a {\it single} entity {\it independent} of selection of an object through their intersection.
Closed loops are desirable here to avoid boundaries, and the need to
specify conditions on them. Subsequently, a sphere, $S^2$,
with $x$ inside it then allows distinction, $y\ne x$, for $y$
outside of $S^2$ (and inside its own sphere). Of course, $S^2$ is curved.
The intersection of two loops locally spans a plane in which the selected object
lies, which constitutes an
implicit selection of two of the three spatial dimensions. For self-consistency, the
selection structure needs to be repeated for the two other combinations of
spatial dimensions, yielding the three-torus, $T^3=S^1\times S^1\times S^1$.
Upto here, an object $x$ should just be viewed as a point.

The distinct objects $x$ and $y$, inside their individual $S^2$ spheres, need to be
connected by a handle, $S^1\times S^2$, embedded in four dimensions.
Because this allows one to use distinct objects {\it themselves} in their
{\it own} selection by looping them through the handle back onto $T^3$, thus rendering the
choice process self-contained globally. This is the crucial
step in the construction of a self-consistent choice structure, and all results below
derive from it. Of course, the presence of handles naturally
incorporates the idea that one can distinguish objects in a spatial and a temporal sense.

Finally, one just needs to consider {\it all} pairs $[x,y]$.
This yields the connected sum of two (one for $x$ and one for $y$) three-tori
and three (the number of possible intersections of two loops on $T^3$) handles.
The resulting structure, $Q$, is a three-dimensional topological
manifold $2T^3\oplus 3(S^1\times S^2)$ embedded in four dimensions, where $\oplus$
denotes the connected sum, using three-ball surgery.
$Q$ can itself select and distinguish all objects that it contains through all
the homeomorphisms (maps that are continuous, are bijections and have continuous inverses)
that it allows.
That is, $Q$ and its homeomorphisms can construct $S^1$ loops and $S^2$ spheres that
facilitate selection and distinction locally.

Any object that is locally chosen through $Q$ can now be
endowed with the symmetries that $Q$ itself possesses. That is, any
object $x$ can also be viewed as a field value ${\bf q}[x]$, provided that the field ${\bf q}$ expresses
only $Q$'s symmetries, i.e., these symmetry properties themselves can {\it also} be selected and
distinguished. This is an expression of the idea that symmetry transformations that $Q$ enjoys
are just manifestations of the ways in which objects in $Q$ are equivalent (as a class).
It is conjectured then that $Q$ is necessary in any self-consistent theory of
fundamental interactions. As such, the guiding principle in this work is that
the mathematical formulation of physical processes should be independent of any
background while the intrinsic nature of these physical processes is, in fact,
determined by the absence of any background.

\section{The Equations of Motion}

To get to dynamical equations, note
first that $T^3$ and $S^1\times S^2$ are prime three-manifolds, i.e., they
cannot be written as the connected sum of other three-manifolds, and thus embody the
notion of discrete, but continuous, building blocks for the choice structure.
Furthermore, a three-torus defines four topologically distinct paths between any two points inside that
three-torus. This is easy to see because an individual three-torus contains three loops, which one can
use to detour over. Also, under homeomorphisms one can
take any path (which is locally $R^3$) and twist it with an element from $U(1)$,
which is just a complex rotation.
One can also exchange the two three-tori in $Q$ without changing the topology.
These properties will be used to derive QED and Einstein gravity later on.

\subsection{General Form}

Now, the interactions and fields present in the equations of motion must follow directly from
the topological structure of $Q$ if one accepts the no background principle.
One is lead then to a single index, denoted by $\lambda$, equation because of the four
topologically distinct paths on $T^3$.
A field ${\bf q}$, whose properties must still follow from the topology
of $Q$, on some path $\lambda$ in $Q$ is selected locally through
another crossing path, say $\mu$. One has a ${\bf q}$ on both of these paths.
The two intersecting $S^1$ loops that define this selection can themselves be
distinguished as two single loops, denoted by case $B$, or
one double loop, denoted by case $F$, on the two three-tori. Clearly, these
two distinctions are
exhaustive in $Q$. One obtains a {\it scalar} ($S^1\times S^1$ is closed) derivative operator on
$[S^1\subset S^2]\times [S^1\subset S^2]$, for case $B$, that is of second order
and is called $D^2$.
Similarly, one finds a scalar first order derivative operator on $[S^1\times S^1]\subset S^2$,
for case $F$, that is called $\delta$.
These operators are still unspecified, but they define ${\bf q}$'s interaction where the $S^1$ loops
cross and select ${\bf q}$.

This ${\bf q}$ interaction must in turn be selected, along any path, by
${\bf q}$ itself, for complete background independence.
This is achieved by looping ${\bf q}$ through a handle, as argued above.
The {\it differences} between these ${\bf q}$ self-interactions, when summed over
all intersecting paths $\mu$, must then be zero for a given path $\lambda$.
This is because the mouths of the
handles can be moved as one likes (under homeomorphisms) from intersection to intersection,
thus nullifying any net distinction between different interactions.
One finds thus, for some path $\lambda$, that
$$q^\mu [q_\lambda D^2 q_\mu -q_\mu D^2 q_\lambda ]=0,\eqno(1B)$$
$$q'^\mu [q'_\lambda\delta q'_\mu -q'_\mu\delta q'_\lambda ]=0.\eqno(1F)$$
For case $B$ and $F$, i.e., for different loop distinctions, ${\bf q}$ cannot be the same object,
which is indicated by a prime, ${\bf q'}$, for case $F$ above. Note that indices refer to paths here and
that, under a homeomorphism, any of the four paths can be chosen to
coincide with any of the four local coordinate directions.
Looking at these simple equations, one might fear that the no background principle is too
limiting, but the solutions to (1) will turn out to be promising.

It is obvious that under $B$ one distinguishes (counts)
all selections that the individual $S^1$ loops define, and there are arbitrarily
many of those, while under $F$ one
distinguishes only one object, the two intersecting $S^1$ loops themselves. Clearly, the
latter case allows a natural interpretation in terms of the Pauli
exclusion principle for fermions.
Case $B$ then refers to bosons.

Indices are raised in (1) through $q^\mu\equiv g^{\mu\nu}q_\nu$ for an, still
unspecified, object ${\bf g}$ on the
four-manifold bounded by $2T^3\oplus 3(S^1\times S^2)$. It is ${\bf g}$ that
measures self-selection by first acting linearly on ${\bf q}$ and ${\bf q'}$ and then
facilitating the contractions in equations (1). Hence, ${\bf g}$ must be {\it symmetric}
since when $x$ selects $y$, the converse is also true.
One has, equally unspecified, covariant (just meaning a lower index) derivatives
$\nabla^{B,F}_\mu$. For case $B$, $D^2\equiv \nabla^{B\mu} \nabla^B_\mu$, and
for $F$, $\delta\equiv i\gamma^\mu \nabla^F_\mu$ with $\gamma^\mu$ the Dirac
matrices. The definition of $\delta$ just follows from the interpretation of case $F$.
In this then, ${\bf q'}$ is a Grassmann, anti-commuting, variable, with ${\bf q'}^2=0$.
This immediately eliminates the second term in equation $(1F)$.

Terms in the covariant derivatives must result from the different symmetries of $Q$
(see QED and Einstein gravity below).
Conversely, for a simply connected region one has locally, i.e., in flat space, that
$g^{\mu\nu}=\eta^{\mu\nu}$, for the Minkowski metric ${\eta^{\mu\nu}}$,
and $\nabla_\nu =\partial_\nu$, provided that the manifold $Q$ supports this metric
signature. The latter is immediate, since it is a statement of special relativity and $Q$,
by construction, is background-free. So information propagating
on $Q$ must do so in the same way, i.e., at the same speed, to render all observers identical.

Finally, further contraction with $q^\lambda$ and $q'^\lambda$
renders the left hand sides of (1) {\it identically} zero, as the
no background principle requires. Any other result would have lead to
an immediate failure of the construct. Also note
then that the {\it form} of equations (1) is determined solely by the topology
of $Q$ and thus is invariant under homeomorphisms.

\subsection{QED}

One sees immediately, for $\nabla^{B,F} =\partial$, that the source-free wave
equations, $D^2 q_\lambda=0$, solve $(1B)$, while the (dimensionless) Dirac equation,
$\delta q'_\lambda =q'_\lambda$, solves (1F). Furthermore, one can
introduce a source field ${\bf s}[{\bf q,q'}]={\bf s}[{\bf q'}]$ to distinguish
$D^2 {\bf q}$ under case $F$. That is, one merely incorporates the fact that $Q$
distinguishes fermions and bosons, and that these can interact.
Also, ${\bf s}$ must be linear in ${\bf q'}$ because selection is a linear action.
The definition $A_\lambda\equiv q_\lambda$ then gives the new solution
$D^2 A_\lambda =s_\lambda$ and the constraint on the source
$$A^\mu A_\lambda s_\mu -A^\mu A_\mu s_\lambda =0.\eqno(2B)$$
Equation $(2B)$ can only be satisfied without the need for an extra symmetry,
which would break self-consistency, only if the properties
of a {\it single} path, say $\mu =\lambda =\lambda_0$, are instrumental in QED.
Of course, ${\bf s}={\bf A}$ is also a solution, but does not express any
interaction between fermions and bosons.
Now, with a slight abuse in notation, allowed because
of form invariance of all equations, the same indices $\lambda$
will also be used to indicate the four {\it coordinate} directions along $\lambda_0$.

Then recall that any path $\lambda_0$ allows a $U(1)$ twist. Of course, this is
just the gauge freedom of $\lambda_0$, denoted as $e^{if}$ for some function $f(x)$.
As a consequence, equation $(1F)$ should be invariant under the
addition of an object to the interaction operator $\nabla^F$.
Obviously, this extra term must be selected by, and be linear in, ${\bf q}$, for self-consistency.
Thus, the modified $\nabla^F_\mu =\partial_\mu +iA_\mu$ enters into (1F).
In this, a single distinct path $\lambda_0$ implies that ${\bf A}$ is coupled
through the first order derivative w.r.t.\ $f$ of $e^{if}$. That is, only the
difference between the boundaries of the path $\lambda_0$ enters into the gauge transformation.
In all, this constitutes the (dimensionless) content of QED.

\subsection{Einstein Gravity}

The interchange symmetry of the two identical three-tori in $Q$,
$T^3\leftrightarrow T^3$, now yields the source object ${\bf T}[{\bf q,q'}]={\bf T'}[{\bf q,q'}]$
for case $B$ {\it and} $F$, given that the loop topologies that express bosons and fermions are both
subsets of an individual three-torus.
Obviously, the interchange symmetry also forces one to consider only {\it pairs}
of paths, denoted for simplicity by ($c_1$,$c_2$), for which $S^1=c_1 c_2^{-1}$ must hold.
That is, as one travels from one three-torus to the other on $Q$, the interchange symmetry
moves one right back to the starting point through identification.
If one first considers case $B$ for the dynamics, then one has
$[{\bf gq}]\rightarrow {\bf R}[{\bf gq}]$ for the {\it change}, along $S^1$, in
${\bf q}$ as one combines all four paths on a three-torus with one another,
i.e., one finds an object $R_{\mu\nu\kappa\lambda}$ with four indices.
So ${\bf R}$ must yield an object $G_{\mu\nu}[{\bf R,g}]$ that is
restricted to $S^1$ and that has only two indices, one for $c_1$ and one for $c_2$.
Equation $(2B)$ is then transformed,
for ${\bf s}\rightarrow {\bf T}$ and ${\bf A}={\bf q}\rightarrow {\bf G}$, to
$$G^{\alpha\beta}G_{\lambda\mu}T_{\alpha\beta}-G^{\alpha\beta}G_{\alpha\beta}T_{\lambda\mu} =0.\eqno(3B)$$
Again, because of form invariance of all equations, the same indices
are used for the coordinate directions $\mu\nu$ along the two individual paths.
On $S^1$ any order along $c_1$ and $c_2$ can be taken, so one has that $T_{\mu\nu}=T_{\nu\mu}$ and
that $G_{\mu\nu}=G_{\nu\mu}$.
One immediately finds from $(3B)$ that $G_{\mu\nu}=T_{\mu\nu}$ must hold along $S^1$.
Note that ${\bf q}$ now implicitly depends on ${\bf G}$ and that ${\bf q}[{\bf G}]={\bf q}[{\bf T}]$
is the quantity that contains all the information about the dynamics. So one should
always transform back to a ${\bf q}$ (or ${\bf q'}$ under $F$) form of the equations of motion
to confirm the validity of the found solution. For case $B$ this is trivial and
the shape (geometry) of $S^1$ is fixed by ${\bf G}={\bf T}$.

For case $B$, ${\bf g}$ can now be found given that ${\bf R}$
must be fixed by the {\it difference} $q^\mu_{c_2}-q^\mu_{c_1}$, i.e., the distinction
between the two paths. So, one has again linearity in ${\bf q}$ and
$$q^\mu_{c_2}-q^\mu_{c_1}=q_0^\kappa R^\mu_{\kappa\lambda\nu}c_1^\lambda c_2^\nu,\eqno(4B)$$
for the displacements $c_1^\lambda ,c_2^\nu$ along ($c_1,c_2$) and an arbitrary
point $0$. As expected, this is just a statement of what curvature, embodied by ${\bf R}$, the
field ${\bf q}$ sees. The maximum
order of derivatives of ${\bf g}$ in ${\bf G}[{\bf R,g}]$ is two because there are two
distinct paths that enter, and ${\bf G}$ is linear in these second derivatives because
the interchange symmetry acts through $S^1$ (and not some higher power thereof).
The constraint $\nabla^{B\mu} G_{\mu\nu}=0$ must hold {\it locally} under
$T^3\leftrightarrow T^3$ since $\nabla^B{\bf G}[c_1]=\nabla^B{\bf G}[c_2]=-\nabla^B{\bf G}[c_1]$.

Also, for case $F$, under $\delta {\bf q'}={\bf s'}\rightarrow {\bf T}$ and ${\bf q'}\rightarrow {\bf G'}$,
the left hand side of equation $(1F)$ transforms analogously to case $B$ and becomes
$$G'^{\alpha\beta}G'_{\lambda\mu}T_{\alpha\beta}=0.\eqno(3F)$$
One has identical results as above since ${\bf G'}={\bf T}$ solves $(3F)$,
after transforming back to ${\bf q'}$ form, through ${\bf q'}[{\bf G'}]{\bf q'}[{\bf T}]=0$.
Again, because $Q$ is self-consistent, the field ${\bf q'}$ that traces the dynamics of space-time
must be used instead of ${\bf G'}$. Hence, fermions and bosons see the same geometry,
${\bf q'}[{\bf G'}]={\bf q'}[{\bf G}]$ for case $F$ and
${\bf q}[{\bf G'}]={\bf q}[{\bf G}]$ for case $B$.

Most importantly, the equivalence principle follows quite naturally from the no background principle
since the only sense of mass, as a {\it distinct} property, that $Q$ possesses is through the curvature
of the handles. So mass is an expression of topology here.
Finally then, ${\bf g}$ describes a spin 2 field, because of the $T^3$ interchange
symmetry, with four dynamic degrees of freedom from the four distinct paths on a three-torus in $Q$.
In all, this constitutes the (dimensionless) content of Einstein gravity.

\section{Discussion}

Obviously, the ease with which the fundamentals of QED and Einstein gravity are reproduced is a
direct consequence of the contractions with $q^\mu$ and $q'^\mu$ that are present
in $(1B)$ and $(1F)$, which in turn is just a statement of self-selection, the heart
of the no background principle.
For gravity alone, the no background principle is then just a reformulation of
Mach's principle, and nothing is new. However, one might argue that $Q$ provides a physical
interpretation for the axiom of choice in that a self-consistent choice
function, as constructed above, actually leads to physical constraints.
In a unified theory, all symmetries that $Q$ allows should be
considered of course. To this effect, it can be shown that $Q$ also contains $SU(N)$
symmetries as well as relative mass scales and boundary
conditions\cite{1,2}. Finally, the presence of the handles that allow the
self-selection of objects, strongly resembles the picture of Wheeler's space-time foam.

\end{document}